%% file: main.tex
\documentclass[11pt,a4paper]{article}

\usepackage[utf8]{inputenc}
\usepackage[english]{babel}
\usepackage{amsmath, amssymb, amsthm}
\usepackage{graphicx}

\usepackage{geometry}
\usepackage{cite}
\usepackage{booktabs}
\usepackage{caption}
\usepackage{subcaption} 
\usepackage{orcidlink}
\usepackage{algorithm}
\usepackage{algorithmicx}
\usepackage{algpseudocode} 

\usepackage{hyperref}

\title{\textbf{A Non-Monotonic Relationship: An Empirical Analysis of Hybrid Quantum Classifiers for Unseen Ransomware Detection}}

\author{
  Huu Phu Le$^{1}$ \\
  \texttt{phule9225@gmail.com,\orcidlink{0009-0000-9668-7756}} \\
  \and
  Phuc Hao Do$^{2,*}$ \\  
  \texttt{do.hf@sut.ru,\orcidlink{0000-0003-0645-0021}} \\
  \and
  Vo Hoang Long Nguyen$^{1}$ \\
  \texttt{longnguyen.080400@gmail.com,\orcidlink{0009-0007-6844-7776}} \\
  \and
  Nang Hung Van Nguyen$^{3}$ \\
  \texttt{nguyenvan@dut.udn.vn,\orcidlink{0000-0002-9963-7006}} \\
}

\date{
    $^{1}$Danang Architecture University, Da Nang, Vietnam \\
    $^{2}$Bonch-Bruevich St. Petersburg State University of Telecommunications, Russia \\
    $^{3}$University of Science and Technology – The University of Danang, Vietnam \\[2ex]
    $^{*}$Corresponding author: do.hf@sut.ru \\[2ex]
}

\begin{document}

\maketitle

\begin{abstract}
Detecting unseen ransomware is a critical cybersecurity challenge where classical machine learning often fails. While Quantum Machine Learning (QML) presents a potential alternative, its application is hindered by the dimensionality gap between classical data and quantum hardware. This paper empirically investigates a hybrid framework using a Variational Quantum Classifier (VQC) interfaced with a high-dimensional dataset via Principal Component Analysis (PCA). Our analysis reveals a dual challenge for practical QML. A significant information bottleneck was evident, as even the best-performing 12-qubit VQC fell short of the classical baseline's 97.7\% recall. Furthermore, a non-monotonic performance trend—where performance degraded when scaling from 4 to 8 qubits before improving at 12 qubits—suggests a severe trainability issue. These findings highlight that unlocking QML's potential requires co-developing more efficient data compression techniques and robust quantum optimization strategies.
\end{abstract}


\section{Introduction}
\label{sec:intro}

The digital landscape is locked in a persistent arms race between cybersecurity defenders and malicious actors. Among the most potent weapons in the modern cybercriminal's arsenal is ransomware, a class of malware that encrypts critical data and extorts victims for its release, causing billions of dollars in economic damage and paralyzing essential services annually \cite{brinkley2024machine}. The primary challenge in combating this threat lies in the rapid, continuous evolution of new ransomware variants. Attackers employ sophisticated obfuscation and polymorphism techniques, rendering traditional signature-based detection methods obsolete. While classical machine learning (ML) has emerged as a powerful defense, capable of identifying known threats with high accuracy, it often struggles with the crucial task of detecting novel, "zero-day" ransomware families. Models trained on existing samples frequently fail to generalize, overfitting to the specific characteristics of seen families and leaving networks vulnerable to the unknown.

In the quest for a new detection paradigm with superior generalization capabilities, Quantum Machine Learning (QML) has emerged as a compelling frontier \cite{martin2022quantum}. At its core, QML leverages the principles of quantum mechanics to process information in ways that are classically intractable. Variational Quantum Classifiers (VQCs), a leading class of hybrid quantum-classical algorithms, are particularly promising. By encoding classical data into high-dimensional quantum Hilbert spaces, VQCs can theoretically access exponentially larger and more complex feature spaces. This enhanced expressive power holds the potential to uncover intricate, non-linear patterns hidden within the data that may be invisible to classical models, thereby offering a path toward more robust generalization for zero-day threat detection.

However, translating this theoretical promise into practical application within the current Noisy Intermediate-Scale Quantum (NISQ) \cite{bharti2022noisy} era is fraught with significant, often underestimated, challenges. This paper focuses on two fundamental hurdles that stand at the classical-quantum interface. The first is the dimensionality chasm: real-world cybersecurity datasets are characterized by thousands of features, whereas today's quantum processors consist of only a few dozen noisy qubits. This necessitates aggressive classical dimensionality reduction, creating a severe \textit{information bottleneck} where critical discriminative information may be irretrievably lost before it ever reaches the quantum circuit. The second hurdle is the \textit{trainability crisis}, most notably the phenomenon of barren plateaus \cite{mcclean2018barren}. As the number of qubits and circuit depth increase, the optimization landscape can become exponentially flat, causing training algorithms to stagnate and rendering the quantum model effectively untrainable.

This paper provides a sober, empirical investigation into the interplay of these two challenges in the context of unseen ransomware detection. We implement a hybrid framework combining Principal Component Analysis (PCA) for dimensionality reduction and a VQC for classification. By systematically comparing this framework against strong classical baselines (Logistic Regression, Random Forest, and XGBoost) across 4, 8, and 12-qubit configurations, we uncover a revealing non-monotonic performance trend \cite{padha2022quantum}. Our results demonstrate that naively scaling quantum resources is not a panacea; in fact, it can degrade performance due to the onset of trainability issues. Our core contribution is a realistic assessment of QML's current readiness for this critical cybersecurity task, highlighting that future progress depends not on isolated advancements in quantum hardware, but on the co-development of information-preserving data encoding techniques and robust quantum optimization strategies.

The remainder of this paper is organized as follows: Section~\ref{sec:related} reviews related work in classical and quantum machine learning for cybersecurity. Section~\ref{sec:method} details our methodology, including the dataset and the design of our experimental framework. Section~\ref{sec:experiments} presents and analyzes the comparative performance results. Section~\ref{sec:discussion} interprets these findings and discusses their broader implications, and Section~\ref{sec:conclusion} concludes the paper with a summary and directions for future research.

\section{Related Work} \label{sec:related}

This research lies at the confluence of machine learning (ML) for cybersecurity \cite{apruzzese2023role} and the rapidly evolving domain of quantum computing. To contextualize our contributions, we review prior work in three critical areas: classical machine learning approaches for ransomware detection, the development and application of Variational Quantum Classifiers (VQCs) \cite{li2022recent} in Quantum Machine Learning (QML), and the theoretical challenges associated with training quantum neural networks, particularly the barren plateau phenomenon. These areas collectively frame the opportunities and limitations of applying QML to the challenging task of detecting unseen ransomware.

\subsection{Classical Machine Learning for Ransomware Detection}
The application of machine learning to ransomware and malware detection has been a cornerstone of modern cybersecurity research. Classical ML approaches have primarily focused on two paradigms: static analysis and dynamic analysis. Static analysis involves extracting features directly from the binary structure of portable executable (PE) files \cite{gibert2022pe} without executing them. Commonly used features include API calls, DLL imports, strings, and byte n-grams, which are computationally efficient to extract and provide a rich representation of a program's structure. 

These features have been successfully leveraged to train a variety of classical models, such as Support Vector Machines (SVMs), Random Forests, and Gradient Boosting \cite{sahin2020assessing} methods like XGBoost. These models have demonstrated high detection rates, often achieving accuracy and recall scores above 95\% on datasets containing known malware families.

In contrast, dynamic analysis executes malware in a controlled sandbox environment to capture behavioral features, such as registry modifications, file system operations, and network communications. This approach is particularly effective against obfuscation techniques \cite{behera2015different}, which are commonly employed by ransomware to evade static analysis. However, dynamic analysis is resource-intensive, requiring significant computational infrastructure and time, making it less practical for real-time detection scenarios. 

\subsection{Variational Quantum Classifiers}
Quantum Machine Learning (QML) has emerged as a promising frontier for addressing complex classification tasks, particularly in scenarios where classical models struggle with generalization. Variational Quantum Classifiers (VQCs), a prominent class of QML algorithms, are designed for the Noisy Intermediate-Scale Quantum (NISQ) era, where quantum hardware is limited in scale and susceptible to noise. VQCs operate as hybrid quantum-classical models, combining a parameterized quantum circuit (PQC) with a classical \cite{sim2019expressibility} optimizer. 

A critical component of VQCs is the feature map, a quantum circuit that encodes classical data \cite{ghosh2021encoding} into a high-dimensional quantum Hilbert space. Havlíček et al. \cite{havlivcek2019supervised} demonstrated that such feature maps can create computationally complex state spaces that are difficult to simulate classically, potentially offering a quantum advantage over classical kernel methods, such as those used in SVMs. For instance, the \texttt{ZZFeatureMap} \cite{singh2025modeling} used in our study leverages entangling gates to capture intricate correlations in the data, which could theoretically enhance generalization for complex tasks. Since their introduction, VQCs have been applied to diverse domains, including particle physics, financial modeling, and medical image classification \cite{amini2016classification}, achieving promising results in controlled settings. However, their application to cybersecurity, particularly ransomware detection, remains underexplored, with only a few studies investigating quantum approaches for malware classification. This gap motivates our work, as we seek to evaluate the practical feasibility of VQCs in a high-stakes, real-world cybersecurity context.

\subsection{Trainability Challenges in Quantum Neural Networks}
Despite their theoretical potential, training VQCs and other quantum neural networks presents significant challenges, particularly in the NISQ era. One of the most formidable obstacles is the \textit{barren plateau} phenomenon, first identified by McClean et al \cite{mcclean2018barren} Barren plateaus occur when the variance of the cost function’s gradient vanishes exponentially as the number of qubits or circuit depth increases. This results in a flat optimization landscape, where classical optimizers, such as gradient-based methods or even gradient-free approaches like COBYLA \cite{pellow2021comparison}, struggle to find productive update directions, rendering the model effectively untrainable. This phenomenon is particularly problematic for deep or highly parameterized quantum circuits, which are often necessary to capture complex data patterns.

Subsequent research has identified additional factors that exacerbate barren plateaus. For example, the choice of measurement strategy can significantly impact trainability, with global observables (e.g., measuring all qubits simultaneously) \cite{mcdermott2005simultaneous} often leading to flatter landscapes compared to local observables. Moreover, noise in quantum hardware, a hallmark of NISQ devices, can induce noise-driven barren plateaus, further complicating optimization. These theoretical insights are critical for interpreting empirical VQC performance \cite{mcdermott2005simultaneous}, as they suggest that increasing the number of qubits or circuit complexity does not necessarily improve results and may, in fact, degrade performance due to optimization failures.

\section{Methodology} \label{sec:method}

The methodology of this study is designed to systematically compare classical and hybrid quantum-classical approaches \cite{do2025challenges} for detecting unseen ransomware, addressing the critical challenge of generalizing to zero-day threats. By evaluating both paradigms on a common dataset, we aim to provide a reproducible framework that quantifies the trade-offs between established classical machine learning techniques and emerging quantum methods. This section describes the dataset, the overarching experimental framework, the classical baseline models, and the hybrid quantum-classical framework, with a focus on the mathematical and algorithmic underpinnings of each component.

\subsection{Dataset}
Our experiments utilize the "Ransomware Combined Structural Feature Dataset" \cite{moreira2024ransomware}, which is specifically curated to evaluate zero-day ransomware detection. This dataset comprises 2675 samples, divided into a training set of 2157 samples and a testing set of 518 samples. A distinguishing feature of this dataset is the complete separation of ransomware families between the training and testing sets: the training set includes 25 ransomware families, while the testing set contains 15 entirely distinct families. This disjoint structure ensures a realistic zero-day detection scenario, where models must generalize to novel threats without relying on shared family characteristics. Each sample is represented by a high-dimensional feature vector $\vec{x} \in \mathbb{R}^{D}$, where the initial dimensionality $D = 1567$ corresponds to static features extracted from portable executable (PE) files, such as API calls, DLL imports, and byte n-grams. To ensure consistency and mitigate scale-related biases, all features are normalized using Scikit-learn's \texttt{StandardScaler}, which is fitted solely on the training data to prevent data leakage.

\subsection{Experimental Framework}
\begin{figure}[h!]
\centering
\includegraphics[width=0.9\textwidth]{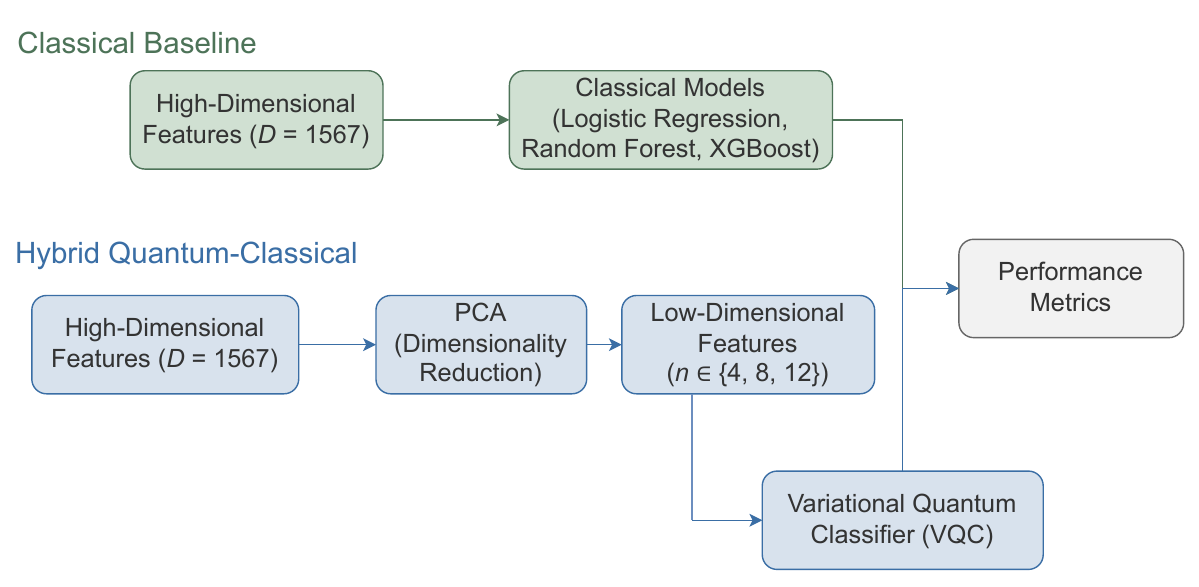} 
\caption{Experimental framework for comparing classical and hybrid quantum-classical models. The classical pipeline (top) processes the full feature set, while the hybrid pipeline (bottom) uses PCA to compress data before feeding it to a VQC.} \label{fig:architecture}
\end{figure}

The experimental framework is structured to facilitate a fair and comprehensive comparison between classical and quantum approaches, as illustrated in Fig.~\ref{fig:architecture}. The framework operates in two parallel stages. The first stage trains and evaluates classical machine learning models on the full high-dimensional feature set ($D = 1567$), establishing a robust performance benchmark. The second stage implements a hybrid quantum-classical pipeline, where the high-dimensional data is compressed into a low-dimensional representation using a classical dimensionality reduction technique, followed by classification using a Variational Quantum Classifier (VQC) \cite{maheshwari2021variational}. By systematically varying the target dimensionality, we investigate the impact of information loss on the quantum model’s performance. This dual-stage approach ensures that both paradigms are evaluated under conditions that reflect their practical constraints, particularly the limited qubit resources in quantum hardware.

The overall process is formalized in Algorithm~\ref{alg:framework}, which outlines the steps for both classical and hybrid stages. The algorithm highlights the iterative nature of the hybrid pipeline, where multiple dimensionality settings are evaluated to assess the trade-off between data compression and classification performance.

\begin{algorithm}[h!]
\caption{Experimental Framework for Classical and Hybrid Quantum-Classical Comparison}
\label{alg:framework}
\footnotesize
\begin{algorithmic}[1]
\State \textbf{Input:} Dataset $X \in \mathbb{R}^{N \times D}$, labels $y \in \{0, 1\}^N$, qubit counts $n \in \{4, 8, 12\}$
\State \textbf{Output:} Performance metrics for classical and quantum models
\State \textbf{Stage 1: Classical Pipeline}
\State Normalize $X$ using \texttt{StandardScaler} fitted on training data
\State Split $X$ into training ($X_{\text{train}}$) and testing ($X_{\text{test}}$) sets
\For{each classical model $M \in \{\text{LR}, \text{RF}, \text{XGBoost}\}$}
    \State Train $M$ on ($X_{\text{train}}, y_{\text{train}}$)
    \State Evaluate $M$ on ($X_{\text{test}}, y_{\text{test}}$) using accuracy, precision, recall, F1-score, AUC
\EndFor
\State \textbf{Stage 2: Hybrid Quantum-Classical Pipeline}
\For{each $n \in \{4, 8, 12\}$}
    \State Compute PCA projection matrix $W_n \in \mathbb{R}^{D \times n}$ using $X_{\text{train}}$
    \State Project data: $X_{\text{pca}} = X W_n$
    \State Split $X_{\text{pca}}$ into training ($X_{\text{pca,train}}$) and testing ($X_{\text{pca,test}}$) sets
    \State Initialize VQC with \texttt{ZZFeatureMap} and \texttt{RealAmplitudes} ansatz
    \State Train VQC on ($X_{\text{pca,train}}, y_{\text{train}}$) using COBYLA optimizer
    \State Evaluate VQC on ($X_{\text{pca,test}}, y_{\text{test}}$) using accuracy, precision, recall, F1-score, AUC
\EndFor
\State \textbf{Return:} Performance metrics for all models
\end{algorithmic}
\end{algorithm}

\subsection{Classical Baseline Models}
To establish a performance benchmark, we employ three classical machine learning algorithms, each selected for its proven efficacy in handling high-dimensional tabular data. Logistic Regression (LR) serves as a simple yet effective linear model, modeling the probability of a sample being ransomware via the logistic function:
\begin{equation}
    p(y = 1 | \vec{x}) = \frac{1}{1 + e^{-\vec{w}^T \vec{x} - b}},
\end{equation}
where $\vec{w} \in \mathbb{R}^D$ and $b \in \mathbb{R}$ are the learned weights and bias, respectively. Random Forest (RF) leverages an ensemble of decision trees, aggregating their predictions to achieve robustness against overfitting. XGBoost, a state-of-the-art gradient boosting framework, optimizes a sequence of decision trees to minimize a regularized loss function, often outperforming other methods on structured data. These models are implemented using their standard configurations in the Scikit-learn and XGBoost libraries, with hyperparameters tuned via cross-validation on the training set to ensure optimal performance.

\subsection{Hybrid Quantum-Classical Framework}
The hybrid quantum-classical framework integrates classical preprocessing with a quantum classification core, addressing the dimensionality mismatch between classical data and quantum hardware. This framework is divided into two components: a classical-quantum interface using Principal Component Analysis (PCA) and a quantum core based on a Variational Quantum Classifier (VQC).

\subsubsection{The Classical-Quantum Interface: PCA}
Given the high dimensionality of the dataset ($D = 1567$) and the limited number of qubits available on current quantum hardware ($n \ll D$), direct encoding of the data into a quantum circuit is infeasible. To address this, we apply PCA, a linear dimensionality reduction technique that projects the data onto a lower-dimensional subspace while preserving maximum variance. For a centered data matrix $X \in \mathbb{R}^{N \times D}$, PCA computes a projection matrix $W_n \in \mathbb{R}^{D \times n}$ by solving the optimization problem:
\begin{equation}
    W_n = \arg\max_{W} \left( \text{Tr}(W^T X^T X W) \right) \quad \text{subject to} \quad W^T W = I_n,
\end{equation}
where $I_n$ is the $n \times n$ identity matrix, and $\text{Tr}(\cdot)$ denotes the trace. The resulting low-dimensional data is obtained as:
\begin{equation}
    X_{\text{pca}} = X W_n, \quad X_{\text{pca}} \in \mathbb{R}^{N \times n}.
\end{equation}
We select $n \in \{4, 8, 12\}$ to align with the qubit constraints of our quantum circuits. The cumulative explained variance, shown in Table~\ref{tab:pca_variance}, quantifies the information retained after projection, revealing significant loss: 19.14\% for 4 components, 29.07\% for 8 components, and 35.50\% for 12 components. This loss underscores the information bottleneck challenge, as a substantial portion of the original data’s variance is discarded.

\begin{table}[h!]
\caption{Cumulative Explained Variance vs. Number of PCA Components.}
\label{tab:pca_variance}
\centering
\begin{tabular}{cc}
\toprule
\textbf{Number of Components (Qubits)} & \textbf{Cumulative Explained Variance (\%)} \\
\midrule
4 & 19.14 \\
8 & 29.07 \\
12 & 35.50 \\
\bottomrule
\end{tabular}
\end{table}

\subsubsection{The Quantum Core: Variational Quantum Classifier}
The VQC is a hybrid quantum-classical algorithm that leverages a parameterized quantum circuit (PQC) for classification, trained using a classical optimizer. For a low-dimensional data point $\vec{x}_{\text{pca}} \in \mathbb{R}^n$, the VQC operates in three stages: state preparation, parameterized transformation, and measurement.

In the state preparation stage, the classical data vector $\vec{x}_{\text{pca}}$ is encoded into an $n$-qubit quantum state using a feature map circuit $\mathcal{U}_{\Phi(\vec{x}_{\text{pca}})}$. We employ the \texttt{ZZFeatureMap}, a second-order Pauli expansion that introduces entanglement through two-qubit interactions, defined as:
\begin{equation}
    \mathcal{U}_{\Phi(\vec{x}_{\text{pca}})} = \prod_{j=1}^n e^{-i x_j Z_j} \prod_{j<k} e^{-i (x_j - x_k)^2 Z_j Z_k},
\end{equation}
where $Z_j$ is the Pauli-Z operator on the $j$-th qubit, and $x_j$ is the $j$-th component of $\vec{x}_{\text{pca}}$. This maps the input to a quantum state:
\begin{equation}
    |\psi(\vec{x}_{\text{pca}})\rangle = \mathcal{U}_{\Phi(\vec{x}_{\text{pca}})} |0\rangle^{\otimes n}.
\end{equation}

In the parameterized transformation stage, a trainable ansatz $U(\vec{\theta})$, parameterized by a vector $\vec{\theta}$, is applied to the encoded state. We use the \texttt{RealAmplitudes} ansatz, a hardware-efficient circuit comprising layers of single-qubit Y-rotations and CNOT gates for entanglement. The final quantum state is:
\begin{equation}
    |\psi_{\text{final}}\rangle = U(\vec{\theta}) \mathcal{U}_{\Phi(\vec{x}_{\text{pca}})} |0\rangle^{\otimes n}.
\end{equation}

In the measurement and prediction stage, an observable $M$, typically a Pauli operator such as $Z_0 \otimes I_{n-1}$, is measured to compute the expectation value:
\begin{equation}
    f(\vec{x}_{\text{pca}}, \vec{\theta}) = \langle \psi_{\text{final}} | M | \psi_{\text{final}} \rangle \in [-1, 1].
\end{equation}
This value is mapped to a binary class prediction (0 or 1) using a threshold (e.g., 0). The VQC is trained by minimizing a cost function, such as the mean squared error between predictions and true labels:
\begin{equation}
    C(\vec{\theta}) = \frac{1}{N_{\text{train}}} \sum_{i=1}^{N_{\text{train}}} \left( f(\vec{x}_{\text{pca},i}, \vec{\theta}) - y_i \right)^2,
\end{equation}
where $N_{\text{train}}$ is the number of training samples, and $y_i \in \{0, 1\}$ is the true label. We use the gradient-free COBYLA optimizer to iteratively update $\vec{\theta}$, balancing computational efficiency with convergence.

This hybrid framework, combining PCA and VQC, enables us to evaluate the feasibility of quantum methods for ransomware detection while highlighting the challenges of information loss and quantum circuit optimization.

\section{Results and Analysis} \label{sec:experiments}

This section details the empirical evaluation of both classical and hybrid quantum-classical models for the task of unseen ransomware detection. We begin by outlining the experimental setup, followed by a presentation of the classical baseline performance, which sets a high benchmark. Subsequently, we analyze the performance of the Variational Quantum Classifier (VQC) models, dissecting the results to reveal the fundamental challenges of information loss and trainability that currently hinder the practical application of Quantum Machine Learning (QML) in this domain.

\subsection{Experimental Setup}
All experiments were conducted in a Python 3.10 environment. Classical models were implemented using the Scikit-learn and XGBoost libraries, while the quantum circuits were designed and simulated using Qiskit v1.4.4. To ensure a fair comparison and isolate the algorithmic performance from hardware noise, all VQC experiments were executed on Qiskit's noise-free \texttt{AerSampler} backend. A global random seed of 42 was used across all stages to ensure the reproducibility of data splits, initializations, and other stochastic processes.

The VQCs were trained using the COBYLA optimizer, a gradient-free method well-suited for the potentially noisy cost landscapes of quantum circuits. The number of iterations was set to 100 for the 4-qubit model and adjusted to 80 for the 8 and 12-qubit models to maintain a reasonable computational runtime while allowing for convergence. The primary evaluation metrics included accuracy, precision, recall, F1-score, and the Area Under the Receiver Operating Characteristic Curve (AUC). Given the critical nature of cybersecurity threats, we prioritize \textbf{recall} as the most important metric, as failing to detect a real ransomware sample (a false negative) has far more severe consequences than misclassifying a benign file (a false positive).

\subsection{Performance of Classical Baselines: A High Benchmark}
The classical machine learning models, trained on the complete 1567-dimensional feature set, established a formidable performance benchmark, demonstrating the effectiveness of traditional methods on this dataset. The detailed results are presented in Table~\ref{tab:results}.

Logistic Regression emerged as the top performer, achieving an outstanding recall of \textbf{97.66\%} and an AUC of 0.9920. Random Forest and XGBoost also delivered exceptional results, with recall scores of 95.84\% and 95.06\%, respectively, and both achieving an AUC of 0.9940. These near-perfect scores indicate that the high-dimensional feature space contains strong, highly discriminative patterns that allow these models to generalize effectively to unseen ransomware families.

\begin{figure}[h!]
    \centering
    \includegraphics[width=0.7\textwidth]{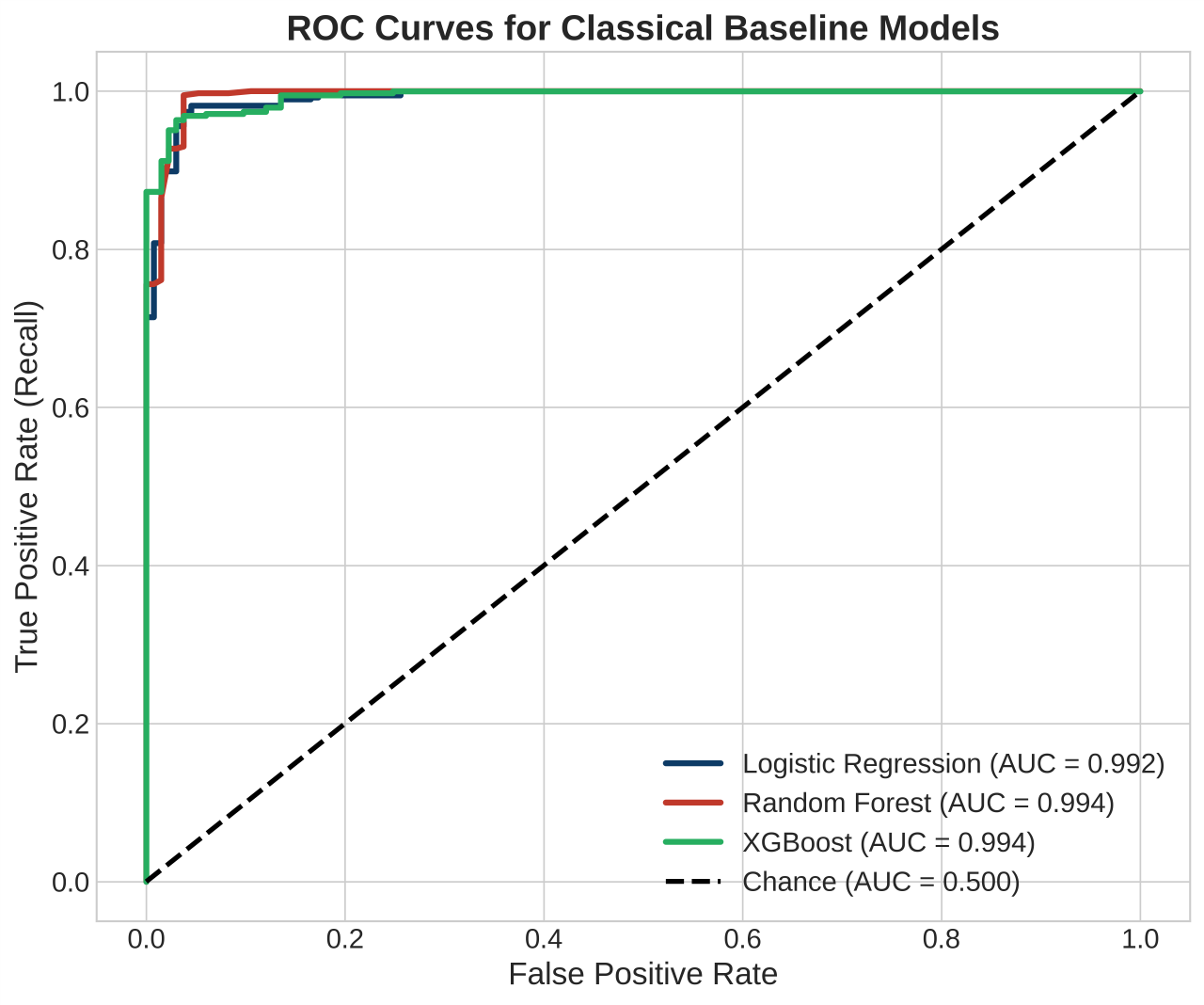}
    \caption{ROC curves for classical models, demonstrating their high discriminative power on the unseen ransomware test set. The curves' proximity to the top-left corner indicates excellent performance across all classification thresholds.}
    \label{fig:roc_classical}
\end{figure}

The ROC curves in Figure~\ref{fig:roc_classical} visually corroborate these strong quantitative results. All three models exhibit curves that hug the top-left corner of the plot, indicative of high true positive rates across various low false positive rates. The near-perfect AUC values, approaching the ideal value of 1.0, signify that the models can reliably distinguish between ransomware and benign samples across all classification thresholds. This exceptional performance by classical algorithms underscores the magnitude of the challenge for any emerging quantum approach to demonstrate practical value.

\subsection{Performance of Hybrid Quantum-Classical Models: A Tale of Two Challenges}
In stark contrast to the classical baselines, the hybrid VQC models, which operated on PCA-reduced data, struggled to achieve competitive performance. As shown in Table~\ref{tab:results}, the performance of all VQC configurations was significantly lower than that of even the simplest classical model. The best-performing 12-qubit VQC only managed a recall of 55.06\%, creating a performance deficit of over 42 percentage points compared to Logistic Regression. This considerable gap points to fundamental obstacles in the hybrid pipeline, which we identify as a dual challenge: a severe information bottleneck and a critical trainability issue.

\begin{table}[h!]
\caption{Performance on Unseen Ransomware Test Set (\% except AUC)}
\label{tab:results}
\footnotesize
\centering
\begin{tabular}{lccccc}
\toprule
\textbf{Model} & \textbf{Accuracy} & \textbf{Precision} & \textbf{Recall} & \textbf{F1-Score} & \textbf{AUC} \\
\midrule
\multicolumn{6}{l}{\textit{Classical Baselines}} \\
Logistic Regression & 97.10 & 98.43 & \textbf{97.66} & 98.04 & 0.9920 \\
Random Forest & 95.95 & 98.66 & 95.84 & 97.23 & 0.9940 \\
XGBoost & 95.75 & 99.19 & 95.06 & 97.08 & 0.9940 \\
\midrule
\multicolumn{6}{l}{\textit{Hybrid VQC Models}} \\
VQC (4 Qubits) & 48.65 & 78.47 & 42.60 & 55.22 & 0.5994 \\
VQC (8 Qubits) & 44.98 & 74.04 & 40.00 & 51.94 & 0.4624 \\
VQC (12 Qubits) & 51.74 & 73.36 & 55.06 & 62.91 & 0.5374 \\
\bottomrule
\end{tabular}
\end{table}

The first challenge, the \textit{information bottleneck}, is a direct consequence of the necessary dimensionality reduction. As detailed in Table~\ref{tab:pca_variance} from our methodology, the PCA preprocessing step results in a substantial loss of information. Even for the 12-qubit case, only 35.50\% of the original data's variance is retained. This means that nearly two-thirds of the statistical information that classical models leveraged so effectively is discarded before the VQC even begins processing. This loss imposes a hard ceiling on the quantum model's potential performance.

\begin{figure}[h!]
    \centering
    \includegraphics[width=0.7\textwidth]{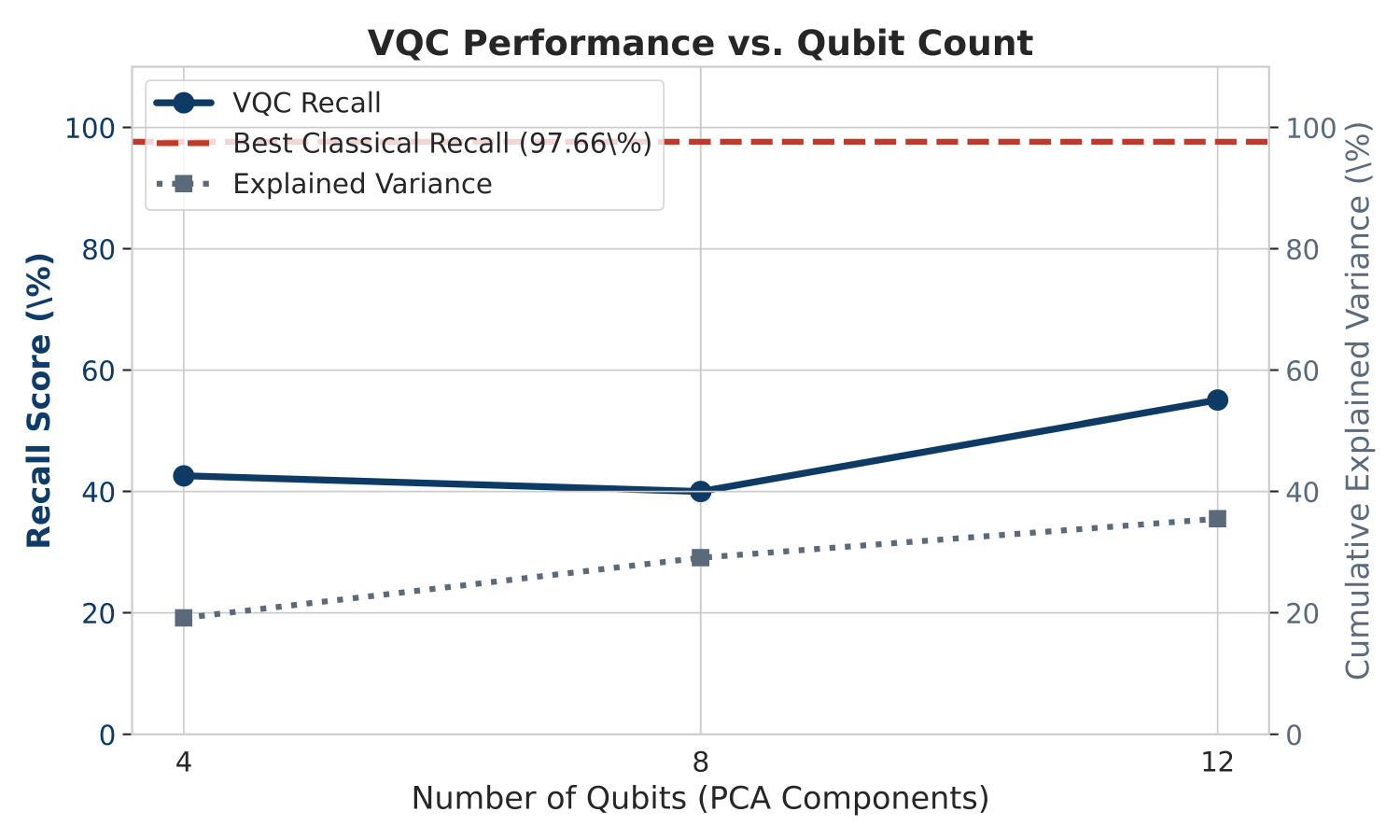}
    \caption{A non-monotonic relationship between VQC recall and the number of qubits. Performance drops when scaling from 4 to 8 qubits before improving at 12 qubits, suggesting a significant trainability challenge.}
    \label{fig:recall_vs_qubits}
\end{figure}

More perplexingly, the models' performance does not scale monotonically with the number of qubits, as illustrated in Figure~\ref{fig:recall_vs_qubits}. While one might expect performance to improve with more quantum resources (and thus more retained information), we observe a distinct performance degradation when moving from 4 qubits (42.60\% recall) to 8 qubits (40.00\% recall). Performance only recovers and surpasses the 4-qubit level at 12 qubits. This counter-intuitive, non-monotonic trend strongly suggests the presence of underlying optimization challenges, which we investigate further in the following section.

\subsection{Unpacking the Trainability Challenge: An Analysis of Training Dynamics}

To diagnose the root cause of the non-monotonic performance and confirm the suspected trainability issues, we inspected the convergence behavior of each VQC model by examining the cost function values recorded during the optimization process. This analysis provides direct insight into how effectively the COBYLA optimizer navigated the parameter landscape of each model, even without a visual plot.

Our analysis of the training logs revealed a clear distinction in optimization behavior across the different model sizes. The cost function for the 4-qubit model consistently decreased throughout the iterations, indicating that the optimization process was making tangible progress and successfully finding a path toward a better solution. 

In stark contrast, the 8-qubit model exhibited immediate stagnation. Its cost function showed negligible improvement from its initial high value, remaining effectively flat for the majority of the training run. This observed lack of convergence is a classic symptom of a barren plateau, a phenomenon where the optimization landscape becomes so flat that the optimizer is effectively lost, unable to find a productive direction of descent. The training dynamics of the 12-qubit model were situated between these two extremes, showing a slight decrease in the cost function but far less pronounced than that of the 4-qubit case, which correlates with its marginal performance improvement.

This evidence from the training process confirms that the performance drop at 8 qubits is not an anomaly but a direct consequence of an optimization failure. As the number of qubits and circuit parameters increased from 4 to 8, the VQC appears to have entered a barren plateau regime, rendering it effectively untrainable with the given optimizer and iteration count. The marginal improvement at 12 qubits suggests that the benefit of retaining more data variance began to slightly overcome the severe training difficulty, but the underlying optimization problem remains a critical bottleneck. This analysis solidifies our conclusion that simply scaling up qubits is not a panacea and can be detrimental without co-evolving optimization strategies to handle increasingly complex and challenging training landscapes.

\section{Discussion} \label{sec:discussion}

This study explores the application of a hybrid quantum-classical framework for detecting unseen ransomware, offering valuable insights into the current capabilities and limitations of Quantum Machine Learning (QML) in cybersecurity. By comparing classical machine learning models with a Variational Quantum Classifier (VQC) combined with Principal Component Analysis (PCA), we highlight the significant challenges that prevent quantum approaches from matching the performance of traditional methods. The discussion reflects on these findings, focusing on the interplay between data compression and optimization difficulties, while the conclusion outlines a path forward for future research to enhance QML’s potential in this critical domain.

\subsection{The Dichotomy Between Classical and Quantum Performance}
The results reveal a clear divide between the performance of classical and quantum models in detecting zero-day ransomware. Classical models, particularly Logistic Regression, excelled with a recall of 97.66\% and an AUC of 0.9920 when trained on the full 1567-dimensional feature set. This impressive performance suggests that the dataset contains strong, distinguishable patterns that allow even a straightforward linear model to effectively identify novel ransomware threats. The success of classical methods establishes a high standard, challenging any new approach to achieve comparable accuracy in real-world cybersecurity scenarios.

In stark contrast, the hybrid VQC-PCA framework struggled to compete, with its best configuration at 12 qubits achieving only 55.06\% recall and an AUC of 0.5374. This significant gap is driven by two key obstacles. First, the need to compress high-dimensional data to fit the limited qubit capacity of quantum circuits creates a substantial loss of information. PCA, while effective at reducing dimensionality, retains only a fraction of the data’s original patterns—35.50\% at 12 qubits—discarding critical details needed to distinguish ransomware from benign samples. The noticeable improvement in recall from 40.00\% at 8 qubits to 55.06\% at 12 qubits shows that the VQC benefits from retaining more information, but the large amount of discarded data still severely limits its ability to generalize to new threats. This underscores a fundamental issue: if the classical preprocessing step eliminates essential information, the quantum circuit cannot recover it, regardless of its potential to process complex patterns.

The second challenge is the difficulty of training the VQC effectively. The unexpected pattern in performance, where the 8-qubit model performed worse than the 4-qubit model (AUC of 0.4624 versus 0.5994) before improving at 12 qubits, points to significant optimization hurdles. This non-monotonic trend suggests that increasing the number of qubits complicates the training process, likely due to the barren plateau phenomenon, where the optimization process struggles to find meaningful improvements. The simpler 4-qubit model, with fewer parameters, navigates a less complex training landscape, achieving a modestly better outcome. However, the 8-qubit model appears to get stuck in an ineffective solution, while the 12-qubit model benefits slightly from additional information but still falls short of classical performance. This indicates that simply scaling up quantum resources is not enough without better strategies to guide the training process.

\subsection{The Dual Challenge: Interplay of Information Bottleneck and Trainability}

Our empirical results do not point to a single point of failure for the hybrid quantum-classical model, but rather to a compounding ``dual challenge'' that arises at the intersection of classical data preprocessing and quantum circuit optimization. This challenge represents a fundamental hurdle for applying QML to high-dimensional, real-world problems like cybersecurity.

\paragraph{The Information Bottleneck as a Performance Ceiling:}
The first component of this challenge is the \textit{information bottleneck} created by aggressive dimensionality reduction. As shown in Table~\ref{tab:pca_variance}, even with 12 principal components, we retain only 35.50\% of the original data's variance. This preprocessing step, while necessary to bridge the gap between classical data and NISQ-era hardware, imposes a hard ceiling on the model's potential performance. A quantum classifier, no matter how expressive or powerful, cannot recover the discriminative information that was discarded before the data ever reached the quantum circuit. This limitation explains why even our best-performing VQC fell drastically short of the classical baseline, which had access to the full, information-rich feature space.

\paragraph{The Trainability Trap of Scaling Qubits:}
The second, more subtle component is the \textit{trainability trap}, evidenced by the non-monotonic performance trend (Fig.~\ref{fig:recall_vs_qubits}). The degradation in performance when scaling from 4 to 8 qubits strongly suggests the onset of a barren plateau. While the 4-qubit model operates in a smaller, more navigable optimization landscape, the 8-qubit model, with its increased parameter space and circuit depth, likely enters a region where gradients vanish, rendering the COBYLA optimizer ineffective. The subsequent improvement at 12 qubits indicates that the benefit of retaining more information (from 29.07\% to 35.50\% variance) began to marginally outweigh the increased training difficulty. However, this non-linear relationship reveals a critical insight: naively adding more qubits is not a guaranteed path to better performance and can, in fact, be detrimental if the optimization landscape becomes untrainable.

\paragraph{The Vicious Cycle of Interaction:}
The true challenge lies in the \textit{vicious cycle} created by the interplay of these two issues. The information bottleneck means the VQC is tasked with learning from a weak, noisy, and incomplete signal. The trainability trap means the quantum model struggles to effectively learn even from this degraded signal. In essence, the model is simultaneously ``starved'' of information and ``incapable'' of digesting what little information it receives. This synergy explains the profoundly poor performance and suggests that solving these problems in isolation is insufficient. A breakthrough in quantum advantage for this class of problems will require co-designing data encoding techniques that are both information-preserving and lead to trainable quantum circuits.

\section{Conclusion}\label{sec:conclusion}
In conclusion, this study provides a clear-eyed evaluation of the hybrid VQC-PCA framework for zero-day ransomware detection, demonstrating that its effectiveness is currently hindered by information loss during data compression and difficulties in training quantum circuits. The strong performance of classical models highlights their reliability for this task, while the VQC’s limitations underscore the early stage of QML in practical applications.

To move QML toward practical utility in cybersecurity, future research should focus on overcoming these challenges through a coordinated approach that improves both data preparation and quantum model design. One key area is developing better methods to compress data while preserving its essential patterns. Techniques like autoencoders or variational autoencoders, which can capture more complex relationships than PCA, could provide quantum circuits with richer inputs, potentially improving their ability to detect novel threats. Another critical focus is enhancing the training process for quantum models. Exploring new circuit designs, such as data-reuploading classifiers or quantum convolutional neural networks, could help maintain stability during training, reducing the risk of optimization failures.

\subsubsection*{Code Availability.}
The source code used to perform the experiments and the results presented in this paper are publicly available on GitHub at: \url{https://github.com/ailabteam/qml-ransomware-project}.


\bibliographystyle{plainnat} 
\input{main.bbl}

\end{document}